# CR-precis: A deterministic summary structure for update data streams

Sumit Ganguly[1] and Anirban Majumder[2]⋆

[1] Indian Institute of Technology, Kanpur
[2] Lucent Technologies, Bangalore

sganguly@cse.iitk.ac.in, manirban@lucent.com

**Abstract.** We present the first deterministic sub-linear space algorithms for a number of fundamental problems over update data streams, such as, (a) point queries, (b) range-sum queries, (c) finding approximate frequent items, (d) finding approximate quantiles, (e) finding approximate hierarchical heavy hitters, (f) estimating inner-products, (g) constructing near-optimal $B$-bucket histograms, (h) estimating entropy of data streams, etc.. We also present new lower bound results for several problems over update data streams.

## 1 Introduction

The data streaming model [2,29] presents a viable computational model for monitoring applications, for example, network monitoring, sensor networks, etc., where data arrives rapidly and continuously and has to be processed in an online fashion using sub-linear space. Some examples of fundamental data streaming primitives include, (a) estimating the frequency of items (point queries) and ranges (range-sum queries), (b) finding approximate frequent items, (c) finding approximate quantiles, (d) finding approximate hierarchical heavy hitters, (e) estimating inner-product, (f) constructing approximately optimal $B$-bucket histograms, (g) estimating entropy, etc..

A data stream is viewed as a sequence of arrivals of the form $(i, v)$, where, $i$ is the identity of an item belonging to the domain $\mathcal{D} = \{0, 1, \ldots, N-1\}$ and $v$ is a non-zero integer that depicts the change in the frequency of $i$. $v \geq 1$ signifies $v$ insertions of the item $i$ and $v \leq -1$ signifies $|v|$ deletions of $i$. The frequency of an item $i$ is denoted by $f_i$ and is defined as the sum of the changes to its frequency since the inception of the stream, that is, $f_i = \sum_{(i,v) \text{ appears in stream}} v$. If $f_i \geq 0$ for all $i$ (i.e., deletions correspond to prior insertions) then the corresponding streaming model is referred to as the strict update streaming model (i.e., Turnstile model [29]). The model where $f_i \lesseqgtr 0$ is called the general update streaming model (i.e., general Turnstile model [29]). The insert-only model refers to data streams with no deletions, that is, $v > 0$. For strict update streams or for insert-only streams, $m$ denotes the sum of frequencies, that is, $m = \sum_{i \in \mathcal{D}} f_i$. For general update streams, $L_1$ denotes the standard norm $L_1 = \sum_{i \in \mathcal{D}} |f_i|$.

⋆ Work done while at IIT Kanpur.

*Prior work on deterministic algorithms over update streams.* Despite the substantive advances in algorithms for data stream processing, there are no deterministic sub-linear space algorithms for a family of fundamental problems in the update streaming models including, estimating the frequency of items and ranges, finding approximate frequent items, finding approximate $\phi$-quantiles, finding approximate hierarchical heavy hitters, constructing approximately optimal $B$-bucket histograms, estimating inner-products, estimating entropy, etc.. Deterministic algorithms are often indispensable in practice. For example, in a marketing scenario where frequent items correspond to subsidized customers, a false negative would correspond to a missed frequent customer, and conversely, in a scenario where frequent items correspond to punishable misuse [24], a false positive results in an innocent victim.

Gasieniec and Muthukrishnan [29] (page 31) briefly outline a data structure, that we use later and will now review. We refer to this structure as the CR-precis structure in this paper (because the Chinese Remainder theorem plays a crucial role in our analysis). The structure is parameterized by a height parameter $k$ and a width parameter $t$. Choose $t$ consecutive prime numbers $k \leq q_1 < q_2 < \ldots < q_t$ and keep a collection of $t$ tables $T_j$, for $j = 1, \ldots, t$, where, $T_j$ has $q_j$ integer counters, numbered from $0, 1, \ldots, q_j - 1$. Each stream update of the form $(i, v)$ is processed as follows.

**for** $j := 1$ **to** $t$ **do** $\{\ T_j[i \mod q_j] := T_j[i \mod q_j] + v\ \}$

Lemma 1 presents the space requirement of CR-precis structure and is implicit in [29](pp. 31). Its proof is given in Appendix A.

**Lemma 1.** *The space requirement of a* CR*-precis structure with height parameter $k \geq 12$ and width parameter $t \geq 1$ is $O(t(t + \frac{k}{\ln k}) \log(t + \frac{k}{\ln k})(\log L_1))$ bits. The time required to process a stream update is $O(t)$ arithmetic operations.* □

Gasieniec and Muthukrishnan use this structure to solve the $k$-set problem, namely, given that there are at most $k$ items with non-zero frequency, identify the items and their frequencies. Let $t = (k-1)\log_k N + 1$. With this choice of $t$, the authors argue that each of the top-$k$ items is isolated in some counter (or group) of a table in the data structure, as follows. If $f_x$ and $f_y$ are each non-zero, then, $x$ and $y$ can collide in at most $\log_k N$ counters. Otherwise, the difference $|x - y| < N$ will be divisible by $\log_k N + 1$ different primes, each larger than $k$. The product of these primes is greater than $k^{\log_k N+1} = kN > N$—a contradiction. The authors state that "doing $\log N$ non-adaptive sub-grouping with each other groups above will solve the problem of identifying the toppers"[1] and then claim that the total space required is $poly(k, \log N)$. We note that the $k$-set problem can be solved using space $O(k \log^2(mN))$ bits for strict update streams and using space $O(k^2 \log^2(mN))$ bits for general update streams [16] using a different technique. The authors do not consider any of the variety of the problems that we consider in this paper, including, estimating the frequency of items and ranges, finding approximate frequent items, finding approximate quantiles, constructing approximately optimal $B$-bucket histograms, estimating inner-product sizes, estimating entropy, etc.

---

[1] Gasieniec and Muthukrishnan state the problem as that of finding the top-$k$ items, called $k$-toppers, in a stream with at most $k$ non-zero frequencies. Hence, each item with non-zero frequency qualifies as a topper.

*Contributions.* We present the first deterministic and sub-linear space algorithms for a set of fundamental problems for update streams, including, estimating the frequency of items and ranges, finding approximate frequent items, finding hierarchical heavy hitters, estimating inner-products of a pair of streams, estimating approximate quantiles, constructing approximately optimal $B$-bucket histograms, estimating entropy, etc.. Gasieniec and Muthukrishnan [29] do not consider any of the above-mentioned problems. We use the the data structure of Gasieniec and Muthukrishnan; however, our novelty lies in the effective analysis of the structure using the Chinese Remainder theorem (and hence we name this structure as the CR-precis structure).

We also present new lower bound results. We show that any algorithm that returns an estimate $\hat{f}_i$ of the frequency $f_i$ of an item $i$ in a strict update stream satisfying $|\hat{f}_i - f_i| \le \frac{m}{s}$ with probability at least $\frac{2}{3}$, requires $\Omega(s(\log m)(\log \frac{N}{s}))$ bits. We also show that over general update streams, the problems of finding approximate frequent items, finding approximate quantiles, estimating the entropy and estimating $k^{th}$ norms $L_k$, require $\Omega(N)$ space.

*Organization.* The remainder of the paper is organized as follows. In Section 2, we define data streaming problems of interest. A more detailed review is given in Appendix B. Section 3 presents the technical results in the paper. Finally, we conclude in Section 4.

## 2  Review

In this section, we briefly review some basic problems over data streams.

The *point query* problem with parameter $s$ is the following: given $i \in \mathcal{D}$, obtain an estimate $\hat{f}_i$ such that $|\hat{f}_i - f_i| \le \frac{L_1}{s}$. For *insert-only streams*, the Misra-Gries algorithm [28], rediscovered and refined in [13,4,25], uses $s \log m$ bits and returns $\hat{f}_i$ such that $f_i \le \hat{f}_i \le f_i + \frac{m}{s}$. The *Lossy Counting* algorithm [26] is also a deterministic point query estimator for insert-only streams that returns an estimate satisfying $f_i \le \hat{f}_i \le f_i + \frac{m}{s}$ using $s \log \frac{m}{s} \log m$ bits. *Sticky Sampling* algorithm [26] extends the *Counting Samples* algorithm [17] to return an estimate satisfying $f_i - \frac{m}{s} \le \hat{f}_i \le f_i$ with probability $1 - \delta$ using space $O(s \log \frac{1}{\delta} \log m)$ bits. For *strict update streams*, the COUNT-MIN sketch algorithm satisfies $f_i \le \hat{f}_i \le f_i + \frac{m}{s}$ with probability $1 - \delta$ using space $O(s \log \frac{1}{\delta} \log m)$ bits. For *general update streams*, the COUNT-MIN sketch algorithm satisfies $|\hat{f}_i - f_i| \le \frac{L_1}{s}$ using the same order of space. The COUNTSKETCH algorithm [7] is applicable for *general update streams* and satisfies $|\hat{f}_i - f_i| \le (F_2^{res}(s)/s)^{1/2} \le \frac{m}{2s}$ with probability $1 - \delta$ using space $O(s \log \frac{1}{\delta} \log m)$, where, $F_2^{res}(s)$ is the sum of the squares of all but the

top-$k$ frequencies in the stream. [4] show that any algorithm that returns $\hat{f}_i$ satisfying $|\hat{f}_i - f_i| \leq \frac{m}{s}$ must use $\Omega(s \log \frac{N}{s})$ bits.

An item $i$ is said to be *frequent* with respect to parameter $s$ provided $|f_i| \geq \frac{L_1}{s}$. Since, finding all and only frequent items requires $\Omega(N)$ space [12,25], research has focused on the following problem of finding $\epsilon$-approximate frequent items, where, $0 < \epsilon < 1$ is a parameter: return all frequent items but do not return any $i$ such that $|f_i| < \frac{(1-\epsilon)L_1}{s}$ [7,12,11,13,17,25,28,26,31]. As reviewed in Appendix B, algorithms for finding frequent items typically use point query estimators and return all items whose estimated frequency exceeds the threshold for frequent items. [10] uses COUNT-MIN sketches for finding frequent items over strict update streams with probability $1 - \delta$ using $O(\frac{s}{\epsilon} \log \frac{s \log(N/s)}{\delta} \log \frac{N}{s} \log m)$ bits. The hierarchical heavy hitters problem [9,10,14,24] is a generalizes the frequent items problems to hierarchical domains (see Appendix B).

Given a *range* $[l, r]$ from the domain $\mathcal{D}$, the range frequency is defined as $f_{[l,r]} = \sum_{x=l}^{r} f_x$. The *range-sum query* problem with parameter $s$ is: given a range $[l, r]$, return an estimate $\hat{f}_{[l,r]}$ such that $|\hat{f}_{[l,r]} - f_{[l,r]}| \leq \frac{m}{s}$. A standard approach is to decompose a given interval as a canonical disjoint sum of at most $2 \log N$ dyadic intervals [20] (See Appendix B). [10] uses COUNT-MIN sketches to estimate range-sums using space $O(s \log \frac{\log N}{\delta} \log N \log m)$ bits and with probability $1 - \delta$. Given $0 \leq \phi \leq 1$ and $j = 1, 2, \ldots, \lceil \phi^{-1} \rceil$, an $\epsilon$-approximate $j^{th}$ $\phi$-quantile is an item $a_j$ such that $(j\phi - \epsilon)m \leq \sum_{i=a_j}^{N-1} f_i \leq (j\phi + \epsilon)m$. The problem has been studied in [10,21,19,27]. For insert-only streams, [21] presents an algorithm requiring space $O((\log \epsilon^{-1}) \log(\epsilon m))$ for insert-only streams. For strict update streams, the problem of finding approximate quantiles can be reduced to that of estimating range sums [19] (See Appendix B). [10] uses COUNT-MIN sketches to find $\epsilon$-approximate $\phi$-quantiles with confidence $1 - \delta$ using space $O(\frac{1}{\epsilon} \log^2 N (\log \frac{\log N}{\phi \delta}))$.

A $B$-bucket histogram $h$ is an $N$-dimensional vector with $B$ interval-value pairs as follows. Divide the domain $\mathcal{D} = \{0, 1, \ldots, N-1\}$ into $B$ non-overlapping intervals, say, $I_1, I_2, \ldots, I_B$. For each interval $I_j$, choose a value $v_j$. Then $h$ is the vector such that for each $i \in \mathcal{D}$, $h_i = v_j$, where $I_j$ is the unique interval containing $i$. The cost of a $B$-bucket histogram $h$ with respect to the frequency vector $f$ is defined as $||f - h|| = \sum_{j=1}^{B} \sum_{i \in I_j} (f_i - v_j)^2$. Let $h^{opt}$ denote an optimal $B$-bucket histogram satisfying $||f - h^{opt}|| = \min_{B\text{-bucket histogram } h} ||f - h||$. The problem is to find a $B$-bucket histogram $\hat{h}$ such that $||f - \hat{h}|| \leq (1+\epsilon) ||f - h^{opt}||$. An algorithm for this problem is presented in a seminal paper [18] using space and time poly $(B, \frac{1}{\epsilon}, \log m, \log N)$ and improved in [22].

Given two streams $R$ and $S$ with item frequency vectors $f$ and $g$ respectively, the *inner product* $f \cdot g$ is defined as $\sum_{i \in \mathcal{D}} f_i \cdot g_i$. The problem is to return an estimate $\hat{P}$ satisfying $|\hat{P} - f \cdot g| \leq \Delta$. The work in [1] presents a space lower bound of $s = \Omega(\frac{m^2}{\Delta})$. Randomized algorithms [1,8,15] match the space lower bound, up to poly-logarithmic factors (See Appendix B.)

The *entropy* of a data stream is defined as $H = \sum_{i \in \mathcal{D}} \frac{|f_i|}{L_1} \log \frac{L_1}{|f_i|}$. It is a measure of the randomness, or, the incompressibility of the stream. The prob-

lem is to return an $\epsilon$-approximate estimate $\hat{H}$ satisfying $|\hat{H} - H| \leq \epsilon H$. For insert-only streams, [6] presents an $\epsilon$-approximate entropy estimator that uses space $O(\frac{1}{\epsilon^2} \log \frac{1}{\delta} \log^3 m)$ bits and also shows an $\Omega(\frac{1}{\epsilon^2 \log(1/\epsilon)})$ space lower bound for estimating entropy. For update streams, [3] presents an $\epsilon$-approximate estimator that requires space $O((\epsilon^{-3} \log^5 m)(\log \epsilon^{-1})(\log \delta^{-1}))$. For $\alpha > 1$, an $\alpha$-approximation for $H$ is an estimate $\hat{H}$ such that $Ha^{-1} \leq \hat{H} \leq Hb$ such that $ab \leq \alpha$. $\alpha$-approximate estimators of $H$ are presented in [23] using $O(N^{\frac{1}{\alpha}} \log N)$ bits and in [5] using $O(\min(m^{2/3}, m^{\frac{4}{\alpha+1}}))$ bits [5].

We note that sub-linear space deterministic algorithms are not known for any of the above-mentioned problems.

## 3 CR-precis structure for update streams

In this section, we use the CR-precis structure to present algorithms for a family of basic problems over update streams.

*An application of the Chinese Remainder Theorem.* Consider a CR-precis structure with height $k$ and width $t$. Fix $x \in \{0, \ldots, N-1\}$. Suppose $J \subset \{1, 2, \ldots, t\}$ such that $|J| \geq \log_k N$. How many items $y$ from the domain $\{0, 1, \ldots, N-1\}$ map to the same bucket as $x$ in each of the tables $T_j$, for $j \in J$? By Chinese Remainder theorem, there is a unique solution in the range $0 \leq y \leq \prod_{j \in J} q_j - 1$ to the equations $x \equiv y \mod q_j$, for each $j \in J$. Since, $\prod_{j \in J} q_j > k^{\log_k N} = N$, it follows that the only solution for $y \in \{0, \ldots, N-1\}$ is $x$. Therefore, for any given $x, y \in \{0, 1, \ldots, N-1\}$ such that $x \neq y$,

$$|\{j \mid y \equiv x \mod q_j \text{ and } 1 \leq j \leq t\}| \leq \log_k N - 1 \ . \tag{1}$$

### 3.1 Algorithms for strict update streams

In this section, we use the CR-precis structure to design algorithms over strict update streams.

*Point Queries.* Consider a CR-precis structure with height $k$ and width $t$. The frequency of $x \in \mathcal{D}$ is estimated as: $\hat{f}_x = \min_{j=1}^{t} T_j[x \mod q_j]$. The accuracy guarantees are given by Lemma 2.

**Lemma 2.** *For $0 \leq x \leq N-1$, $0 \leq \hat{f}_x - f_x \leq \frac{(\log_k N - 1)}{t}(m - f_x)$.*

*Proof.* Clearly, $T_j[x \mod q_j] \geq f_x$. Therefore, $\hat{f}_x \geq f_x$. Further,

$$t\hat{f}_x \leq \sum_{j=1}^{t} T_j[x \mod q_j] = tf_x + \sum_{j=1}^{t} \sum_{\substack{y \neq x \\ y \equiv x \mod q_j}} f_y \ .$$

Thus, $t(\hat{f}_x - f_x) = \sum_{j=1}^{t} \sum_{\substack{y \neq x \\ y \equiv x \bmod q_j}} f_y = \sum_{y \neq x} \sum_{j: y \equiv x \bmod q_j} f_y$

$= \sum_{y \neq x} f_y |\{j : y \equiv x \bmod q_j\}| \leq (\log_k N - 1)(m - f_x)$, by (1) . □

If we let $k = s$ and $t = s \log_s N$, then, the space requirement of the point query estimator is $O(s^2 (\log_s N)^2 (\log m))$ bits. The time required to obtain the estimate is $O(t) = O(s \log_s N)$ arithmetic operations. A slightly improved guarantee that is often useful for the point query estimator is given by Lemma 3. Here, $m^{res}(s)$ is the sum of all but the top-$k$ frequencies [3,7].

**Lemma 3.** *Consider a* CR*-precis structure with height $s$ and width $2s \log_s N$. Then, for any $0 \leq x \leq N - 1$, $0 \leq \hat{f}_x \leq \frac{m^{res}(s)}{s}$.*

*Proof.* Let $y_1, y_2, \ldots, y_s$ denote the items with the top-$s$ frequencies in the stream (with ties broken arbitrarily). By (1), $x$ conflicts with each $y_j \neq x$ in at most $\log_s N$ buckets. Hence, the total number of buckets at which $x$ conflicts with any of the top-$s$ frequent items is at most $s \log_s N$. Thus there are at least $t - s \log_s N$ tables where, $x$ does not conflict with any of the top-$s$ frequencies. Applying the proof of Lemma 2 to only these set of $t - s \log_s N \geq s \log_s N$ tables, the role of $m$ is replaced by $m^{res}(s)$. This proves the lemma. □

As reviewed in Section 2 and Appendix B, the problems of estimating range-sums, finding approximate frequent items, finding approximate hierarchical heavy hitters and $\epsilon$-approximate quantiles essentially reduce to point query estimators associated with simple hierarchical data structures (e.g., dyadic interval hierarchy). Further, in the robust $B$-bucket histogram structure of [18], the role of sketches can be replaced by CR-precis structure. Theorem 4 states the space versus accuracy guarantees for these problems over strict update streams. In addition, the structure can be used to deterministically obtain approximate top-$k$ wavelet coefficients and fourier transform coefficients over update streams—we omit the details for brevity.

**Theorem 4. 1.** *There exists a deterministic algorithm for finding $\epsilon$-approximate frequent items with parameter $s$ using space $O(\frac{s^2}{\epsilon^2}(\log_{\frac{s}{\epsilon}} N)(\log \frac{s}{\epsilon}) \log \frac{N}{s} (\log m))$. The time taken to process each stream update is $O(\frac{s}{\epsilon}(\log_{\frac{s}{\epsilon}} N) \log N)$ arithmetic operations.*
**2.** *There exists a deterministic algorithm for range-sum query estimator with parameter $s$ using space $O(s^2(\log_s^2 N) (\log s + \log \log_s N)) (\log m) \log N)$ bits. The time required for processing a stream update is $O(s(\log_s N) (\log N))$ arithmetic operations.*
**3.** *For $\epsilon < \phi$, $\epsilon$-approximate $\phi$-quantiles may be deterministically computed using space $O(\frac{1}{\epsilon^2}(\log^5 N)(\log m)(\log \log N + \log \frac{1}{\epsilon})^{-1})$ bits. The time taken for processing a stream update is $O(\frac{1}{\epsilon}(\log^2 N)(\log \log N + \log \frac{1}{\epsilon})^{-1})$ and for finding each quantile is $O(\frac{1}{\epsilon}(\log^2 N)(\log \frac{1}{\epsilon})(\log \log N + \log \frac{1}{\epsilon})^{-1})$.*

4. *There exists a deterministic algorithm for finding $\epsilon$-approximate hierarchical heavy hitters using space $O(\epsilon^{-2} s^4 h^2 \log(\frac{s^2 h}{\epsilon}) \log m)$ bits where, $h$ is the height of the hierarchy.*
5. *There exists a deterministic algorithm for constructing $(1-\epsilon)$-optimal $B$-bucket histograms using space poly $(B, \frac{1}{\epsilon}, \log m, \log N)$.* □

*Estimating inner product and join sizes.* Let $m_R = \sum_{i \in \mathcal{D}} f_i$ and let $m_S = \sum_{i \in \mathcal{D}} g_i$. We maintain a CR-precis for each of the streams $R$ and $S$, that have the same height $k$, same width $t$ and use the same prime numbers as the table sizes. For $j = 1, 2, \ldots, t$, let $T_j$ and $U_j$ respectively denote the tables maintained for streams $R$ and $S$ corresponding to the prime $q_j$ respectively. The estimate $\hat{P}$ for the inner product is calculated as $\hat{P} = \min_{j=1}^{t} \sum_{b=1}^{q_j} T_j[b] U_j[b]$.

**Lemma 5.** $f \cdot g \leq \hat{P} \leq f \cdot g + \left(\frac{\log_k N}{t}\right) m_R m_S.$

*Proof.* For $j = 1, \ldots, t$, $\sum_{b=0}^{q_j-1} T_j[b] U_j[b] \geq \sum_{b=0}^{q_j-1} \sum_{x \equiv b \bmod q_j} f_x g_x = f \cdot g$. Thus, $\hat{P} \geq f \cdot g$. Further,

$$t\hat{P} \leq \sum_{j=1}^{t} \sum_{b=1}^{q_j} T_j[b] U_j[b] = t(f \cdot g) + \sum_{j=1}^{t} \sum_{\substack{x \neq y \\ x \equiv y \bmod q_j}} f_x g_y$$

$$= t(f \cdot g) + \sum_{x,y: x \neq y} f_x g_y \sum_{j: x \equiv y \bmod q_j} 1$$

$$\leq t(f \cdot g) + (\log_k N - 1)(m_R m_S - f \cdot g), \text{ by } (1). \quad \square$$

Since $f \cdot g$ can be thought of as the size of the natural join of the streams $R$ and $S$ (i.e., $|R \bowtie S|$), this shows that $|R \bowtie S|$ can be approximated up to additive error of $\frac{m_R m_S}{s}$ using $O(s^2 (\log_s^2 N)(\log s)(\log(m)))$ bits.

*Estimating entropy.* A deterministic algorithm that returns an $\alpha$-approximation of $H$ can be designed as follows. We maintain a CR-precis structure of height $k \geq 2$ and width $t = \frac{2m^{1/\alpha}}{\varepsilon \epsilon}$, where, $\alpha, \epsilon$ and $\varepsilon$ are parameters. We first use the *point queries* estimator to find all items $x$ with $f_x \geq \frac{m}{\varepsilon t}$ with additive error of $0 \leq \hat{f}_x - f_x \leq \frac{(m - f_x)}{\varepsilon t}$. Therefore, $f_x \geq (\hat{f}_x - \frac{\varepsilon m}{t})(1 - \frac{\varepsilon}{t})^{-1} = f'_x$ (say). The estimated contribution to entropy by the frequent items is calculated as $\hat{H}_d = \sum_{x: f'_x > \frac{2m}{t}} \frac{f'_x}{m} \log \frac{m}{f'_x}$. Next, we remove the estimated contribution of the frequent items from the tables as follows.

$T_j[i \bmod q_j] := T_j[i \bmod q_j] - f'_i$, for each $i$ s.t. $\hat{f}_i \geq \frac{m}{c}$ and $j = 1, \ldots, t.$

$\hat{H}_s$ estimates the contribution to $H$ by the non-frequent items as follows. $\hat{H}_s = \frac{1}{t} \sum \left\{ T_j[b] \log \frac{m}{T_j[b]} \mid 1 \leq b \leq q_j \text{ and } T_j[b] \leq \frac{m}{\epsilon c} \right\}$. The estimate for $H$ is returned as $\hat{H} = \hat{H}_d + \hat{H}_s$. The space versus accuracy guarantees of the algorithm are summarized in the following lemma.

**Lemma 6.** *For $0 < \varepsilon, \epsilon < \frac{1}{4}$ and $\alpha > 1$, There exists a deterministic algorithm that returns an estimate $\hat{H}$ satisfying $\frac{\hat{H}(1-\varepsilon)}{\alpha} \leq H \leq (1+\epsilon)H$ using space $O(\frac{1}{\epsilon^2} m^{\frac{2(1-\varepsilon)}{\alpha}} (\log^4 m + \log^4 N))$ bits.* □

The proof of Lemma 6 basically uses equation (1) and is omitted for brevity. For insert-only streams, an improvement can be obtained by using algorithm *Frequent* [28,13,25] instead of CR-precis to find the frequent items (and then reducing the CR-precis as before). This gives an $\alpha$-approximation to $H$ using space $O(\frac{1}{\epsilon^2} m^{\frac{1-\varepsilon}{\alpha}} (\log^4 m + \log^4 N))$ and matches the space complexity of the earlier randomized schemes of [5,23], up to poly-logarithmic factors.

*Lower bound.* A standard result [4] shows that any point query estimator with error at most $\frac{m}{s}$ requires $\Omega(s \log \frac{N}{s})$ bits. For strict update streams, we show stronger lower bounds in Lemmas 7 and Lemma 8.

**Lemma 7.** *For $s < \frac{\sqrt{N}}{8}$, any deterministic algorithm that satisfies $|\hat{f}_i - f_i| \leq \frac{m}{s}$ for any $i \in \mathcal{D}$ over strict update streams requires $\Omega(s(\log m) \log \frac{N}{s})$ space.*

*Proof.* Consider a stream consisting of $s^2$ distinct items, organized into $s$ levels with $s$ items per level. The frequency of an item at level $l$ is set to $t_l = \lfloor \frac{2^l}{s} \rfloor$. Let $m_l$ denote the sum of the frequencies of the items in levels 1 through $l$. Let $s' = 8s$. We apply the algorithm $\mathcal{A}(s')$ to obtain the identities of the items, level by level. At iteration $r$, where, $r = 1, \ldots, s$ in succession, we maintain the invariant that items in levels higher than $s - r + 1$ have been discovered and their (exact) frequencies are deleted from the current stream. Let $l = s - r + 1$. By the invariant, at the beginning of iteration $r$, the frequencies are organized into levels 1 through $l$ and $m = m_l$. At iteration $r$, we return the set of items whose estimated frequencies according to $\mathcal{A}(s')$ is at least $t_l - \frac{m_l}{s'}$. Thus, all items at level $l$ are returned. Further, it can be argued that the estimated frequencies of the other items do not cross $t_l$ as follows. We have $m_l = \sum_{l'=1}^{l} s \cdot \lfloor \frac{2^{l'}}{s} \rfloor < 2^{l+1}$. Therefore, $t_l - t_{l-1} = \frac{2^{l-2}}{s} > \frac{2m}{s'}$. At iteration $r$, the items at level $s - r + 1$ are found and their frequencies are deducted. In this manner, after $s$ iterations, the level by level arrangement of the items can be reconstructed. The number of such arrangements is $\binom{N}{s\ s\ldots\ s}$, where, the $s$'s are repeated $s$ times in the multinomial coefficient. Thus, $\mathcal{A}(s')$ requires space $\log \binom{N}{s\ s\ldots\ s} = \Omega(s^2 \log \frac{N}{s})$, since, $N > 64s^2$. Since $s' = 8s$, we have that $\mathcal{A}(8s)$ requires $\Omega(s^2 \log \frac{N}{s})$ bits. The space required is $\Omega(s^2 \log \frac{N}{s}) = \Omega(s(\log m) \log \frac{N}{s})$. This proves the claim for deterministic algorithms. □

**Lemma 8.** *For $s < \frac{\sqrt{N}}{8}$, any randomized algorithm that satisfies $|\hat{f}_i - f_i| \leq \frac{m}{s}$ with probability at least $\frac{2}{3}$ over strict update streams requires $\Omega(s(\log m)(\log \frac{N}{s}))$ bits.*

*Proof.* Consider the bit-vector indexing problem, where, the input is a bit vector $v$ of size $n$ that is presented in full, followed by an index $i$ between 1 and $n$. The problem is to decide whether $v[i] = 1$ or not. This problem requires space $\Omega(n)$ by any randomized algorithm that gives the correct answer with probability $\frac{2}{3}$. We can solve the bit-vector indexing problem with $n = \lfloor s^2 \log \frac{N}{s} \rfloor$ using a point query estimator.

For simplicity, let $s$ divide $N$. A *segment* $\tau$ of $\log N$ indices starting at index 1 mod $\log \frac{N}{s}$, that is, $\tau = a \log \frac{N}{s} + 1, \ldots, (a+1) \log \frac{N}{s}$, is mapped to a pair $(\lambda_\tau, l_\tau)$, where, $\lambda_\tau \in \{a+1, \ldots, (a+1)Ns\}$ and $l_\tau \in \{1, 2, \ldots, s\}$. The mapping $\lambda_\tau$ is defined as follows. First we map the set $S_\tau = \{j - a \log \frac{N}{s} \mid j \in \tau \text{ and } v[j] = 1\}$ to a number $\nu_\tau$ between 0 and $\frac{N}{s}$. Clearly, there are $2^{\log \frac{N}{s}} = \frac{N}{s}$ possibilities for $S_\tau$. $\lambda_\tau$ is a $\log(Ns)$ bit number whose bit representation is $a \circ \nu_\tau$, that is, the higher order $2 \log s$ bits of $\lambda_\tau$ are those of $a$ and the lower order $\log \frac{N}{s}$ bits are those of $\nu_\tau$. The level of $\tau$ is $l_\tau$ and is the $\log s$-bit number $a_{2\log s} a_{2\log s - 2} \cdots a_2$ where $a$ is the $2 \log s$ bit number $a = a_{2\log s} a_{2\log s - 1} \cdots a_1$. Finally, the frequency of $\lambda_\tau$ is set to $f_{\lambda_\tau} = 2^{l_\tau}$. Since each $l_\tau$ is a $\log s$-bit number, there are $s$ levels. Since, $l_\tau = a_{2\log s} a_{2\log s - 2} \cdots a_2$, the number of $\tau$'s with the same value of $l_\tau$ is the number of possible combinations of the odd bit positions of $a$, that is, $a_{2\log s - 1}, a_{2\log s - 3}, \ldots, a_1$. Since, there are $\log s$ such positions, the number of segments $\tau$ with the same value of $l_\tau$ is exactly $2^{\log s} = s$. Moreover, from the construction, it follows that the mapping of segments $\tau$ to pairs $(\lambda_\tau, l_\tau)$ is 1-1, onto and efficiently constructible by storing only one segment at a time.

If the error probability of the point estimator is at most $1 - \frac{1}{3s^2}$, then, it follows using the argument of Lemma 7 that all $j$ with $v[j] = 1$ are retrieved with total error probability bounded by $\frac{s^2}{3s^2} = \frac{1}{3}$. Given a point query estimator that satisfies $|\hat{f}_i - f_i| \leq \frac{m}{s}$ with probability $\frac{2}{3}$, by returning the median of $O(\log s)$ independent estimators boosts the confidence to $1 - \frac{s^2}{3}$. Hence, the space complexity is $\Omega(\frac{s}{\log s}(\log m)(\log \frac{N}{s}))$.

The above argument can be slightly improved as follows. For each $i$, there exists many permutations of the domain $1, \ldots, s^2 \log \frac{N}{s}$ such that the query index $i$ is contained in the segment $\tau$ that is mapped to the highest level $s$. In this configuration, if the point query estimator is invoked to obtain an estimate of $\hat{f}_\tau$, then, by the argument of Lemma 7, $f_\tau$ is completely predicted and therefore, it can be correctly inferred as to whether $v[i]$ is 1 or not. Hence, the space complexity is $\Omega(s(\log m)(\log \frac{N}{s}))$. □

### 3.2 General update streaming model

In this section, we consider the general update streaming model. Lemma 9 summarizes the point query estimator for general update streams.

**Lemma 9.** *Given a CR-precis structure with height $k$ and width $t$. For $x \in \mathcal{D}$, let $\hat{f}_x = \frac{1}{t} \sum_{j=1}^{t} T_j[x \mod q_j]$. Then, $|\hat{f}_x - f_x| \leq \frac{(\log_k N - 1)}{t}(L_1 - |f_x|)$.*

*Proof.* $t\hat{f}_x = \sum_{j=1}^{t} T_j\,[x \mod q_j] = tf_x + \sum_{j=1}^{t} \sum \{f_y \mid y \neq x \text{ and } y \equiv x \mod q_j\}$.

Thus, $t|\hat{f}_x - f_x| = |\sum_{j=1}^{t} \sum_{\substack{y \neq x \\ y \equiv x \mod q_j}} f_y| = |\sum_{y \neq x} \sum_{j: y \equiv x \mod q_j} f_y|$

$\leq \sum_{y \neq x} \sum_{j: y \equiv x \mod q_j} |f_y| \leq (\log_k N - 1)(F_1 - |f_x|)$, by (1) □

Similarly, we can obtain an estimator for the inner-product of streams $R$ and $S$. Let $L_1(R)$ and $L_1(S)$ be the $L_1$ norms of streams $R$ and $S$ respectively.

**Lemma 10.** *Consider a CR-precis structure of height $k$ and width $t$. Let $\hat{P} = \frac{1}{t}\sum_{j=1}^{t}\sum_{b=1}^{t} T_j[b]U_j[b]$. Then, $|\hat{P} - f \cdot g| \leq \frac{(\log_k N - 1)}{t} L_1(R) L_2(S)$.* □

**Lemma 11.** *Deterministic algorithms for the following problems in the general update streaming model requires $\Omega(N)$ bits: (1) finding $\epsilon$-approximate frequent items with parameter $s$ for any $\epsilon < \frac{1}{2}$, (2) finding $\epsilon$-approximate $\phi$-quantiles for any $\epsilon < \phi/2$, (3) estimating the $k^{th}$ norm $L_k = (\sum_{i=0}^{N-1}|f_i|^k)^{1/k}$, for any real value of $k$, to within any multiplicative approximation factor, and (4) estimating entropy to within any multiplicative approximation factor.*

*Proof.* Consider a family $\mathcal{F}$ of sets of size $\frac{N}{2}$ elements each such that the intersection between any two sets of the family does not exceed $\frac{N}{8}$. It can be shown[2] that there exist such families of size $2^{\Omega(N)}$. Corresponding to each set $S$ in the family, we construct a stream $str(S)$ such that $f_i = 1$ if $i \in S$ and $f_i = 0$, otherwise. Denote by $str_1 \circ str_2$ the stream where the updates of stream $str_2$ follow the updates of stream $str_1$ in sequence. Let $\mathcal{A}$ be a deterministic frequent items algorithm. Suppose that after processing two distinct sets $S$ and $T$ from $\mathcal{F}$, the same memory pattern of $\mathcal{A}$'s store results. Let $\Delta$ be a stream of deletions that deletes all but $\frac{s}{2}$ items from $str(S)$. Since, $L_1(str(S) \circ \Delta) = \frac{s}{2}$, all remaining $\frac{s}{2}$ items are found as frequent items. Further, $L_1(str(T) \circ \Delta) \geq \frac{N}{2} - \frac{s}{2}$, since, $|S \cap T| \leq \frac{N}{8}$. If $s < \frac{N}{3}$, then, $\frac{F_1}{s} > 1$, and therefore, none of the items qualify as frequent. Since, $str(S)$ and $str(T)$ are mapped to the same bit pattern, so are $str(S) \circ \Delta$ and $str(T) \circ \Delta$. Thus $\mathcal{A}$ makes an error in reporting frequent items in at least one of the two latter streams. Therefore, $\mathcal{A}$ must assign distinct bit patterns to each $str(S)$, for $S \in \mathcal{F}$. Since, $|\mathcal{F}| = 2^{\Omega(N)}$, $\mathcal{A}$ requires $\Omega(\log(|\mathcal{F}|)) = \Omega(N)$ bits, proving part (1) of the lemma.

Let $S$ and $T$ be sets from $\mathcal{F}$ such that $str(S)$ and $str(T)$ result in the same memory pattern of a quantile algorithm $\mathcal{Q}$. Let $\Delta$ be a stream that deletes all items from $S$ and then adds item 0 with frequency $f_0 = 1$ to the stream. Now all quantiles of $str(S) \circ \Delta = 0$. $str(T) \circ \Delta$ has at least $\frac{7N}{8}$ distinct items, each with frequency 1. Thus, for every $\phi < \frac{1}{2}$ and $\epsilon \leq \frac{\phi}{2}$ the $k$th $\phi$ quantile of the two streams are different by at least $k\phi N$. Part (3) is proved by letting $\Delta$ be an update stream that deletes all elements from $str(S)$. Then, $L_k(str(S) \circ \Delta) = 0$ and $L_k(str(T) \circ \Delta) = \Omega(N^{1/k})$.

---
[2] Number of sets that are within a distance of $\frac{N}{8}$ from a given set of size $\frac{N}{2}$ is $\sum_{r=0}^{\frac{N}{8}} \binom{N/2}{r}^2 \leq 2\binom{N/2}{N/8}^2$. Therefore, $|\mathcal{F}| \geq \frac{\binom{N}{N/2}}{2\binom{N/2}{N/8}^2} \geq \frac{2^{N/2}}{2(3e)^{N/8}} = \frac{1}{2}\left(\frac{16}{3e}\right)^{N/8}$.

Proceeding as above, suppose $\Delta$ is an update stream that deletes all but one element from $str(S)$. Then, $H(str(S) \circ \Delta) = 0$. $str(T) \circ \Delta$ has $\Omega(N)$ elements and therefore $H(str(T) \circ \Delta) = \log N + \Theta(1)$. The multiplicative gap $\log N : 0$ is arbitrarily large—this proves part (4) of the lemma. □

## 4 Conclusions

We present the first deterministic sub-linear space algorithms for a number of fundamental problems over update data streams, including, point queries, range-sum queries, finding approximate frequent items, finding approximate quantiles, finding approximate hierarchical heavy hitters, estimating inner-products, constructing near-optimal $B$-bucket histograms, estimating entropy of data streams, estimating entropy, etc.. We also present new lower bound results for several problems over update data streams.

## A  Proof of Lemma 1

*Proof.* Consider a CR-precis structure of height $k \geq 6$ and width $t$. Denote the $n^{th}$ prime by $p_n$. By Rosser's theorem [30], $p_n \leq n(\ln n + \log n)$, for $n \geq 6$. It follows that if $a = \frac{k}{\ln k + \log k}$, then, $p_a \leq a(\ln a + \log a) < k$. Letting $c = 1 + \ln 2$, we have,

$$\sum_{j=1}^{t} q_j < \sum_{n=a}^{a+t} p_n \leq \sum_{n=a}^{a+t} cn \ln n \leq cp_{a+t} + c \int_{x=a}^{a+t} x \ln x \ dx \leq cp_{a+t} + c \left( \frac{x^2 \ln x}{2} - \frac{x^2}{2} \right) \bigg|_{a}^{a+t}$$

Simplifying the *RHS*, we obtain the statement of the lemma. □

## B  Review

In this Appendix, we present some more details of the basic techniques used in data stream processing with emphasis on processing update streams.

*Preliminaries.* A *dyadic interval* at level $l$ is an interval of size $2^l$ from the family of intervals $\{[i2^l, (i+1)2^l - 1], 0 \le i \le \lceil \frac{N}{2^l} \rceil - 1\}$, for $0 \le l \le \log N$, assuming that $N$ is a power of 2. The set of dyadic intervals of levels 0 through $\log N$ form a complete binary tree as follows. The root of the tree is the single dyadic interval $[0, N-1]$. The nodes at distance $h$ from the root are the set of dyadic intervals at level $\log N - h$. Moreover, for $0 \le h < \log N$, each dyadic interval at level $h$ is of the form $I_h = [i\frac{N}{2^h}, (i+1)\frac{N}{2^h} - 1]$ and has two children at level $h - 1$, namely, the left and the right halves of $I_h$. The left child of $I_h$ is the interval $[2i\frac{N}{2^{h+1}}, (2i+1)\frac{N}{2^{h+1}} - 1]$ and the right child is the interval $[(2i+1)\frac{N}{2^{h+1}}, (2i+2)\frac{N}{2^{h+1}}]$.

Point query estimators can either make one-sided errors or two-sided errors. Point estimators with one-sided errors are either over-estimators, that is, $f_i \le \hat{f}_i \le f_i + \frac{m}{s}$ (for e.g., COUNT-MIN sketch [10]), or, under-estimators, that is, $f_i - \frac{m}{s} \le \hat{f}_i \le f_i$ (e.g., *Counting Samples* [17], *Lossy Counting* [26]). Point estimators with two-sided errors return estimates satisfying $|\hat{f}_i - f_i| \le \frac{m}{s}$ (for e.g., COUNTSKETCH [7]). Algorithms for finding $\epsilon$-approximate frequent items with parameter $s$ typically use point query estimators with parameter $s' = \frac{s}{\epsilon}$. For example, using a one-sided over-estimator, one can return all items $i$ such that $\hat{f}_i \ge \frac{m}{s}$. Estimators with two sided errors can be used to return all items $i$ such that $\hat{f}_i \ge f_i - \frac{m}{s'}$. The problem of efficiently finding $\epsilon$-approximate frequent items can be solved by keeping a point query estimator corresponding to each dyadic level $l = 0, \ldots, \log \frac{N}{s}$ [10]. By construction, each item $i$ belongs to a unique dyadic interval at level $l$, namely, the $l^{th}$ level ancestor of the interval $[i, i]$ in the dyadic tree. The "items" at level $l$ are the set of dyadic intervals $\{[j2^l, (j+1)2^l - 1]\}_{0 \le j \le 2^{d-l}}$ and are identifiable with the domain $\{0, 1, \ldots, 2^{d-l}\}$. With this interpretation, an arrival over the stream of the form $(i, v)$ is processed as follows: update the item $((i \% 2^l), v)$ for each level $l = 0, 1, \ldots, \lfloor \log \frac{N}{s} \rfloor$. The frequency of a dyadic interval $I$ is defined as the sum of the individual frequencies of items in $I$, and is denoted as $f_I$. Since each level 0 item belongs to one and only one dyadic interval at a given level $l$, the sum of the interval frequencies at level $l$ is the same as the sum of the item frequencies at level 0, which is $m$. If an item $i$ is frequent (i.e., $f_i \ge \frac{m}{s}$), then the dyadic interval that contains $i$ at any level $l$ has frequency at least $f_i$ and is therefore also frequent at level $l$. Hence, at each level $l$ starting from $\lfloor \log \frac{N}{s} \rfloor$ and decrementing down to 1, it suffices to consider only those dyadic intervals that are frequent at level $l$. The procedure begins by enumerating $O(s)$ dyadic intervals at level $\lfloor \log \frac{N}{s} \rfloor$ and keeping as candidate intervals whose estimated frequency is at least $\frac{m}{s}$. In general, at level $l$, there are $O(s)$ candidate intervals. For each candidate interval at level $l$, we consider its left and right child intervals at level $l-1$, and repeat the procedure. Since, at any level, the number of candidate intervals is $O(s)$, the total number of intervals considered in the iterations is $O(s \log \frac{N}{s})$. Using the COUNT-MIN sketch algorithm at each dyadic level with total space $O(\frac{s}{\epsilon} \log \frac{s \log(N/s)}{\delta} \log \frac{N}{s})$ counters, one can return all frequent items with probability 1 and not return any item with frequency $\frac{(1-\epsilon)m}{s}$ with probability $1-\delta$. No sub-linear space deterministic algorithms are known for update streams.

The hierarchical heavy hitters problem [9,14] is a useful generalization of the frequent items problem for domains that have a natural hierarchy (e.g., domain of IP addresses). Given a hierarchy, the frequency of a node $X$ is defined as the sum of the frequencies of the leaf nodes (i.e., items) in the sub-tree rooted at $X$. The definition of hierarchical heavy hitter node ($HHH$) is inductive: a leaf node $x$ is an $HHH$ node provided $f_x > \frac{m}{s}$. An internal node is an $HHH$ node provided that its frequency, after discounting the frequency of all its descendant $HHH$ nodes, is at least $\frac{m}{s}$. The problem is, (a) to find all nodes that are $HHH$ nodes, and (b) to not output any node whose frequency, after discounting the frequencies of descendant $HHH$ nodes, is below $\frac{(1-\epsilon)m}{s}$. This problem has been studied in [9,10,14,24]. As shown in [9], it can be solved by using a simple bottom-up traversal of the hierarchy, identifying the frequent items at each level, and then subtracting the estimates of the frequent items at a level from the estimated frequency of its parent [9]. Using COUNT-MIN sketch, the space complexity is $O(\frac{s^2}{\epsilon}(\log \frac{s \log N}{\delta})(\log N)(\log m))$ bits. [24] presents an $\Omega(s^2)$ space lower bound for this problem. Deterministic algorithms for finding HHH items over update streams are not known.

The range-sum query problem, that is, estimating the frequency of a given range, can be solved by using the technique of dyadic intervals [20]. Any range can be uniquely decomposed into the disjoint union of at most $2 \log N$ dyadic intervals of maximum size (for example, over the domain $\{0, \ldots, 15\}$, the interval $[3, 12] = [3, 3] + [4, 7] + [8, 11] + [12, 12]$). The technique is to keep a point query estimator corresponding to each dyadic level $l = 0, 1, \ldots, \log N - 1$. The range-sum query is estimated as the sum of the estimates of the frequencies of each of the constituent maximal dyadic intervals of the given range. Using COUNT-MIN sketch at each level, this can be accomplished using space $O(s \log \frac{\log N}{\delta} \log N \log m)$ bits with probability $1-\delta$ [10]. The problem of finding $\epsilon$-approximate $\phi$-quantiles can be reduced to range-sum queries as follows. For each $k = 1, 2, \ldots, \phi^{-1}$, a binary search is performed over the domain to find an item $a_k$ such that the range sum $f_{[a_k, N-1]}$ lies between $(k\phi - \epsilon)m$ and $(k\phi+\epsilon)m$. A technique for constructing and maintaining $(1-\epsilon)$-*optimal B-bucket histograms* over strict update streams is presented in [18] using space and time $poly\ (B, \frac{1}{\epsilon}, \log m, \log N)$ and improved in [22].

For estimating the inner-product $f \cdot g$, [1] presents the product of sketches technique using space $O(s \log \frac{1}{\delta})$ counters with additive error of $O(\frac{1}{\sqrt{s}}(F_2(R)F_2(S))^{1/2})$, where, $F_2(R) = \sum_{i \in \mathcal{D}} f_i^2$ and $F_2(S) = \sum_{i \in \mathcal{D}} g_i^2$. [1] also presents a space lower bound of $s = \Omega(m^2/(f \cdot g))$ for estimating $f \cdot g$. COUNT-MIN sketches [10] can be used to return an estimate that has additive error of $m^2/s$ with probability

$1 - \delta$ using space $O(s \log \frac{1}{\delta})$. The product of sketches algorithm is improved in [15] to match the space lower bound.